\documentclass[aps, prl, twocolumn]{revtex4-1}

\usepackage{amsmath}
\usepackage{xcolor}
\usepackage{bm}
\usepackage{amsfonts}
\usepackage{verbatim}
\usepackage{parskip}
 \usepackage{graphicx}

\def\bs{\begin{subequations}}
\def\es{\end{subequations}}
\def\aa{\begin{align}}
\def\ab{\end{align}}
\def\ba{\begin{eqnarray}}
\def\ea{\end{eqnarray}}
\def\be{\begin{equation}}
\def\ee{\end{equation}}

\def\ben{\begin{enumerate}}
\def\een{\end{enumerate}}
\def\bs{\bigskip}

\begin{document}

\title{Statistics and Topology of Fluctuating Ribbons}

\author{Ee Hou Yong$^{1,*}$, Farisan Dary$^{1}$, Luca Giomi$^{2}$ and L. Mahadevan$^{3,4,5,\dagger}$  }

\affiliation{$^1$ Division of Physics and Applied Physics, School of Physical and Mathematical Sciences, Nanyang Technological University, Singapore 637371\\
$^2$ Instituut-Lorentz, Universiteit Leiden, P.O. Box 9506, 2300 RA Leiden, Netherlands\\
$^3$ School of Engineering and Applied Sciences, Harvard University, Cambridge, MA 02138\\
$^4$ Department of Physics, Harvard University, Cambridge, MA 02138\\
$^5$ Department of Organismic and Evolutionary Biology, Harvard University, Cambridge, MA 02138
}

\email{eehou@ntu.edu.sg; $^\dagger$ lmahadev@g.harvard.edu\\}

\begin{abstract}

Ribbons are a class of slender structures whose length, width, and thickness are widely separated from each other. This scale separation gives a ribbon unusual mechanical properties in athermal macroscopic settings, e.g. it can bend without twisting, but cannot twist without bending. Given the ubiquity of ribbon-like biopolymers in biology and chemistry, here we study the statistical mechanics of microscopic inextensible, fluctuating ribbons loaded by forces and torques. We show that these ribbons exhibit a range of topologically and geometrically complex morphologies exemplified by three phases - a twist-dominated helical phase (HT), a writhe-dominated helical phase (HW), and an entangled phase - that arise as the applied torque and force is varied. Furthermore, the transition from HW to HT phases is characterized by the spontaneous breaking of parity symmetry and the disappearance of perversions that characterize chirality reversals. This leads to a universal response curve of a topological quantity, the link, as a function of the applied torque that is similar to magnetization curves in second-order phase transitions.  
\end{abstract}



\maketitle


Filamentous structures are ubiquitous in molecular and cellular biology, polymer chemistry and physics \cite{Pillips, deGenne, rubinstein}. These structures are characterized by their geometrical scale separation, whereby one length scale (the length) is very large compared to the other two (the two principal radii), with consequences for their mechanical properties, e.g. they are easy to bend and twist and hard to stretch and shear. In addition, when the smallest of these scales is comparable to a characteristic length $\ell \sim (k_B T/G)^{1/3}$ obtained by balancing the energy associated with thermal fluctuations $k_BT$ ($k_B$ is the Boltzmann constant, and $T$ is the absolute temperature), and enthalpic elasticity $G \ell^3$ ($G$ is the shear modulus of the material), thermally driven fluctuations become important enough to require a statistical treatment of their behavior in the context of physical polymers \cite{deGenne, rubinstein}, biopolymers such as DNA \cite{bust, bust03, strick96} etc. For filamentous objects that are relatively stiff, the worm-like chain (WLC) model \cite{kratky-porod,markosM,markosP, Maha2018}, with an energy that takes the form that is quadratic in the local curvature, has explained a range of experimental observations of the elasticity of DNA \cite{bust,bust03,marko,marko2} and similar biopolymers. Variations that have generalized the original worm-like chain model also account for twisting and stretching deformations \cite{moroz1,moroz2}, locally bistable behavior \cite{levine} etc. and go even further in explaining new observations of slender filaments in passive and active settings.  

However, many biopolymers such as $\alpha$ helices, $\beta$ sheet, graphene nano-ribbon, molybdenum ribbons \cite{yang2013, wang2018, brouhard2018, dogterom2019, zhang2013, ho2002} are ribbon-like, with their cross-sections better described by elongated rectangles rather than circles, and thus require a more sophisticated description beyond the WLC model that account for the anisotropic nature of the cross-section, such as the railway-track model that couples two worm-like chains via transverse bonds \cite{eve,liv,gole}. An alternative is to consider a continuum framework for ribbons, defined as slender structures whose thickness ($h$), width ($w$) and length ($\ell$) are all widely separated, i.e. $h \ll w \ll \ell$. In this setting, deformations of the ribbon will be almost isometric to the Euclidean plane; in the asymptotic limit of zero thickness, its mechanical energy density is solely a function of its mean curvature $H$, following symmetry considerations. At leading order the energy of the ribbon can then be written as $\frac{1}{2} B \int H^2 dA$, where $B=\frac{Eh^{3}}{12(1-\nu^{2})}$ is the bending rigidity of the sheet (made of a material with Young's modulus $E$ and Poisson ratio $\nu$). Defining a ribbon's center-line unambiguously allows us to parametrize the mean curvature of the surface locally in terms of its conical generators \cite{Wunderlich}, and thence effectively integrate the energy in the width-wise direction and reduce it to a one-dimensional theory. An additional simplification arises when further assumes that deformations are small enough so that the edge of regression of the ribbon generated from its center-line does not intersect the physical ribbon, and defines the Sadowsky ribbon; this leads to an energy that is only a function of the curvature and torsion of the center-line \cite{sad}.

The Sadowsky ribbon is isometric to a flat strip at all temperatures and encodes a non-trivial interaction between the local bend and twist degrees of freedom. 
At an experiential level, the reader is invited to cut a long ribbon from a sheet of paper and convince herself of the asymmetry in the bend-twist coupling inherent in these objects: a ribbon can be bent without twisting, but cannot be twisted without bending. This is very different from the behavior of slender filaments with a circular cross-section, where the local twist is completely independent of the geometric torsion of the center-line \cite{love}, although there is a global relation between the two objects via a topological relation \cite{white1}. While there has been increasing interest in the zero-temperature limit of the Sadowsky ribbon starting about two decades ago \cite{Maha1993,staro,Neukirch2021}, the role of finite temperature driven fluctuations on the morphology of ribbons remains essentially unstudied, with one exception \cite{lucamaha} that studied the statistical mechanics of a free ribbon.  
Here, we build on this observation to study the statistical mechanics of the Sadowsky ribbon under the effects of external forces and torques and understand the statistical morphology of fluctuating ribbons by characterizing its geometry and topology as a function of an appropriately scaled temperature and external loads. 

\section*{Geometry, topology and elasticity of a ribbon}
\label{model}

\subsection*{Geometry and topology of a ribbon}

\begin{figure}[htb!]
\centering
   \includegraphics[width=.5\textwidth]{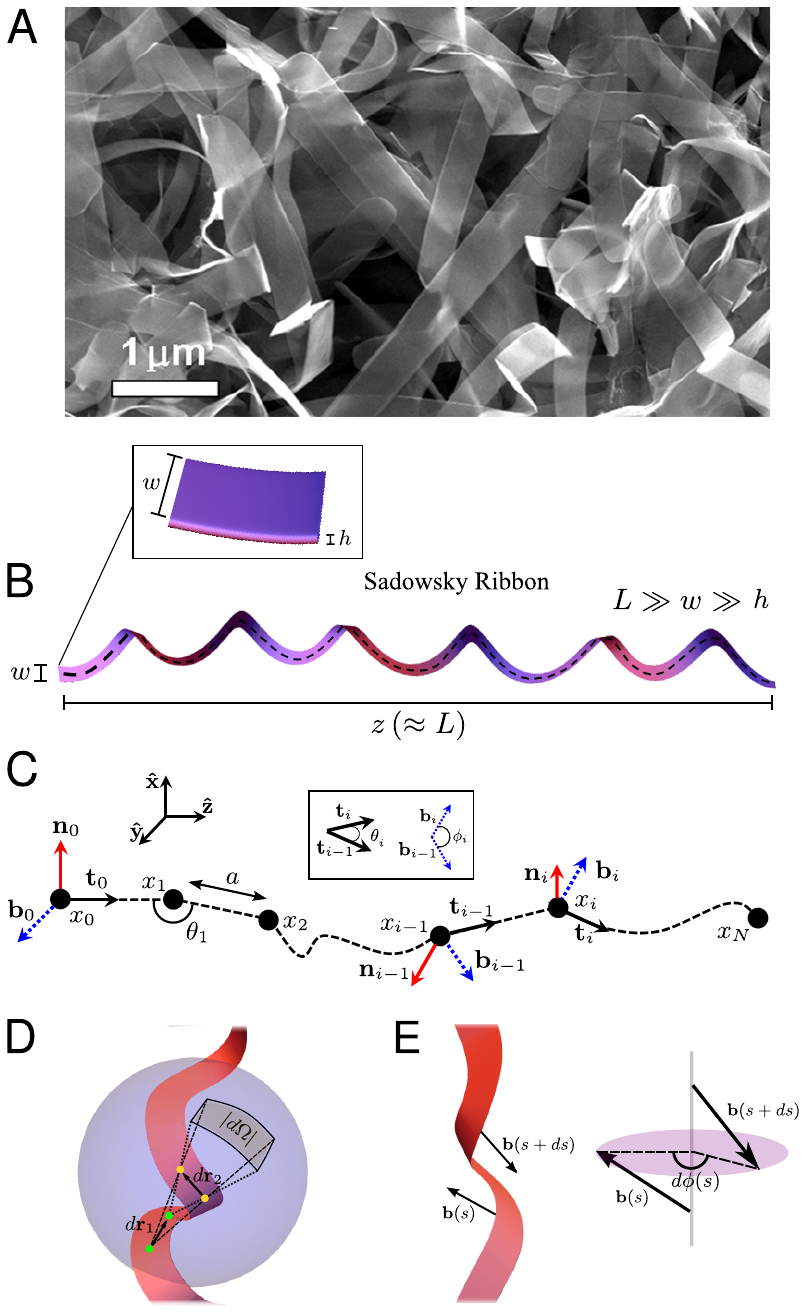}
    \caption{(A) Graphene-coated ribbons of vanadium oxide. Image courtesy of Ref.~\cite{yang2013}. (B) The Sadowsky ribbon with relative extension $z$. Typical dimensions of the ribbon: $h \sim 1$nm, $a \approx 10$ nm, $w/a \approx 1$, $L \sim 10\mu$m, hence $L \gg w \gg h$. The smooth ribbon is generated using {\it BSplineFunction} in Mathematica. (C) A discrete ribbon model consist of $N+1$ vertices, $\mathbf{x}_{0},\mathbf{x}_{1},\, ...\,,\mathbf{x}_{N}$ and an orthogonal Frenet frame ${\bm F}_i = (\mathbf{t}_i, \mathbf{n}_i, \mathbf{b}_i)$ at each vertex. The angle between $\mathbf{t}_{i-1}$ and $\mathbf{t}_{i}$ is $\theta_{i}$; the angle between $\mathbf{b}_{i-1}$ and $\mathbf{b}_{i}$ is $\phi_{i}$; the distance between adjacent vertices is $|\mathbf{x}_{i+1}-\mathbf{x}_{i}|=a$. (D) Writhe, Wr, is the Gauss double integral of the solid angle $d\Omega$ determined by the crossing of $d\mathbf{r}_1$ and $d\mathbf{r}_2$ about each other. (E) Twist, Tw, is the integral of the angle of rotation $d\phi$ of the vector $\mathbf{b}(s)$ about the center-line of the ribbon. }
    \label{Fig:model}
\end{figure}

In the asymptotic limit of a slender ribbon with $h \ll w \ll L$, we can define its configuration completely in terms of the curvature $\kappa(s)$ and torsion $\tau(s)$ as a function of the arc-length $s$ along its center-line as shown in Fig.~\ref{Fig:model}B. 
To understand the interplay between the geometry and topology of the ribbon, we recall the C\u alug\u areanu-White-Fuller theorem Lk = Tw + Wr \cite{white1, fuller1, fuller2} which connects the number of times the two edges of the ribbon whirl around each other, described by the link (also known as linking number), Lk, a global quantity, to the sum of the integrated spatial rate of cross-section rotation, the total twist, Tw, and a global configurational integral that describes the non-planarity of the configuration in terms of the writhe, Wr. For a fixed linking number, the conformations of the ribbon dictate how Lk is distributed between the degrees of freedom associated with Tw and Wr.


\subsection*{Continuum elastic theory} 

Modeling the ribbon as an achiral and inextensible elastic rectangular strip with $h \ll w \ll L$, and the (asymptotic) assumption of developability, 
leads to the Sadowsky energy functional \cite{sad,lucamaha, staro, corrSF1,corrSF2}:
\be
    E_{\text{Sadowsky}} = \frac{1}{2}Bw\int_{0}^{L}ds\, \frac{\left(\kappa^{2}(s)+\tau^{2}(s)\right)^{2}}{\kappa^{2}(s)}.
    \label{Eq:Sadowsky}
\ee
We note that the functional above reduces to that for the planar {\em Elastica}, when $\tau = 0$, but is quite different from the energy functional for a three-dimensional elastic filament that would have an additional term  quadratic in the twisting strain (which is not the geometric torsion of the center-line in general). However, for isometric deformations of a ribbon, the geometric torsion is exactly equal to the local twist since the Frenet frame associated with the center-line coincides with the material frame attached to the cross-section \cite{Maha1993}.
The torsional and flexural deformations are nonlinearly (and asymmetrically) coupled such that the ribbon can store curvature when untwisted (i.e. $\kappa \ne 0$, $\tau =0$) but it becomes energetically prohibitive to store torsion in the presence of a vanishing curvature (i.e. $\kappa \approx 0$, $\tau \ne 0$). 

In the presence of applied end forces and torques, we assume that one end of the ribbon is anchored (fixed boundary) and the other end experiences an applied force $\mathbf{F} = F\hat{\bf z}$ ($F>0$) and a torque $\Omega$, resulting in two additional terms to the free energy:
\begin{align}
    E_{\text{force}} &= -{\bf F} \cdot \int_0^L \mathbf{t}(s) ds  = - Fz, \label{Eq:Eforce} \\
    E_{\text{torque}} &= -2\pi \Omega \text{Lk}, \label{Eq:Etorque}
\end{align}
where $z$ is the end-to-end extension of the ribbon along $\hat{\bf z}$ \cite{markosM,markosP} and
$\text{Lk}$ is the link of the open ribbon \cite{Maha2019}. 
The applied torque acts as a chemical potential for Lk and couples the bend and twist fluctuations, similar to earlier approaches \cite{moroz1,moroz2} where the desired link is achieved by tuning the applied torque on the ribbon. 
Then the complete energy functional for the ribbon is given by
\be
E_{\text{ribbon}}  = E_{\text{Sadowsky}} + E_{\text{force}} + E_{\text{torque}}.
\label{Eq:TotalE}
\ee
We note that our formulation neglects steric effects so that the ribbon is not self-avoiding. 

\subsection*{Discrete model of ribbon} 

In order to understand the implications of the theory for the conformational phase space of the ribbons, we discretize the ribbon into a chain of $N$ plaquettes with $N+1$ vertices $\{\mathbf{x}_{0},\mathbf{x}_{1},\, ...\,,\mathbf{x}_{N}\}$
separated by a fixed length $a$ as shown in Fig. \ref{Fig:model}C, where $a$ is the size of each plaquette and $L = Na$.
For each pair of nearest neighbor vertices $\mathbf{x}_{i-1}$ and $\mathbf{x}_i$, we define the unit tangent vector as $\mathbf{t}_i = (\mathbf{x}_{i} - \mathbf{x}_{i-1})/|\mathbf{x}_{i} - \mathbf{x}_{i-1}|$, for $i = 1,\dots, N$. Next, we define the discrete bending angle $\theta_i$ by $\cos \theta_i =  \mathbf{t}_i\cdot \mathbf{t}_{i-1}$ and the discrete bond angle $\phi_i$ by $\cos \phi_i =  \mathbf{b}_i\cdot \mathbf{b}_{i-1}$.  
At each vertex, there is an orthonormal discrete Frenet frame ${\bm F}_i = (\mathbf{t}_i, \mathbf{n}_i, \mathbf{b}_i)$ as shown in Fig.~\ref{Fig:model}C. The Frenet frames ${\bm F}_i$ are orthogonal $3 \times 3$ matrices, whose column vectors are $\mathbf{t}_i, \mathbf{n}_i$ and $\mathbf{b}_i$. 
Each discrete plaquette is defined by $\mathbf{t}_i$ and $\mathbf{b}_i$. We can write down an iterative relation between two adjacent discrete Frenet frame \cite{lucamaha, Quine04}: ${\bm F}_i =  {\bm F}_{i-1} {\bm R}_i$, where
\begin{equation}
{\bm R}_i = \begin{pmatrix}
\cos\theta_i & -\sin\theta_i &  0 \\[5pt]
\sin\theta_i \cos\phi_i & \cos\theta_i\cos\phi_i &  -\sin\phi_i  \\[5pt]
\sin\theta_i \sin\phi_i  & \cos\theta_i\sin \phi_i &\cos\phi_i 
\end{pmatrix}.
\end{equation}
The discrete curvature at vertex $i$ can be calculated using $\kappa^{2}_{i}=a^{-2} |\mathbf{t}_{i}-\mathbf{t}_{i-1}|^{2} = 2a^{-2}(1-\cos\theta_{i})$. Similarly, the discrete torsion is given by $\tau^{2}_{i}=a^{-2}|\mathbf{b}_{i}-\mathbf{b}_{i-1}|^{2} = 2a^{-2}(1-\cos\phi_{i})$.
The discretized version of the elastic energy functional described by Eq.~\ref{Eq:Sadowsky} to \ref{Eq:TotalE} is
\begin{align}
\begin{split}
    \frac{E_{\text{ribbon}}}{k_{B}T}=&\frac{Bw}{ak_{B}T}\sum_{i=1}^{N-1}\frac{\left[(1-\cos\theta_{i})+(1-\cos\phi_{i})\right]^{2}}{(1-\cos\theta_{i})} \\
     & -\frac{Fa}{k_{B}T}\sum_{i=0}^{N-1}\text{t}^{z}_{i} -\frac{2\pi \Omega}{k_{B}T}\text{Lk},
 \end{split}
 \label{Eq:DiscreteE}
 \end{align}
where $\text{t}^{z}_{i}$ is the $z$-component of the $i$-th tangent vector, $k_B$ is the Boltzmann constant, and $T$ is the temperature.

To track the geometry and topology of the ribbon, we also define discrete analogs of the Tw via a cumulative twist density function as \cite{klenin}
\be
T(n) = \frac{1}{2\pi}\sum_{i=1}^{n-1}\arccos(\mathbf{b}_{i-1}\cdot \mathbf{b}_{i})\,\text{sign}[\mathbf{b}_{i-1}\cdot \mathbf{t}_{i}],
\label{Eq:Tn2}
\ee
where the positive direction of rotation is defined by the right-hand rule, i.e. $\text{sign}[(\mathbf{b}_{i-1}\times \mathbf{b}_{i})\cdot \mathbf{t}_{i-1}]$. The cumulative twist density function specifies how each segment of the ribbon contributes to the overall twist as we move from one end of the ribbon to the other end. When $n = N$, the cumulative twist density function is identical to the twist, i.e. $T(N) =$ Tw. 
Similarly, we define the cumulative writhe density function as 
\be
W(n)=\frac{1}{2\pi}\sum_{i=2}^{n-1}\sum_{j<i} \Omega_{ij},
\label{Eq:Wn}
\ee
where $\Omega_{ij}$ is the Gauss integral along the segments $a\mathbf{t}_{i}$ and $a\mathbf{t}_{j}$, and calculated according to the protocol in ~\cite{klenin}. When $n = N$, the cumulative writhe density function is identical to the writhe, i.e. $W(N)=$ Wr. Finally, the cumulative link density function is
\be
L(n) = T(n) + W(n),
\ee
and $L(N) =$ Lk. The cumulative density functions thus inform us how each segment of the ribbon contributes to the overall link, twist, and writhe.

\subsection*{Parameter values} 
Biopolymers typically have Young's modulus $E \approx 100$ MPa, $h \approx 1$ nm, and $F\sim 0.01-10\, k_{B}T/\text{nm}$ \cite{bust, markosM, ouyang}. 
For our ribbonlike polymer model, $a \approx 10$ nm, $w/a \approx 1$, $Bw/k_{B}T\approx 10$ nm. 
It is useful to define new dimensionless parameters, $\Lambda$ (normalized temperature), $f$ (normalized force), and $\Gamma$ (normalized torque) as follows:
\begin{align}
    \Lambda = \frac{ak_{B}T}{Bw},\quad f = \frac{Fa^{2}}{Bw},\quad \Gamma = \frac{\Omega a}{Bw}. 
\end{align} 
They can be mapped to the original parameters via $f/\Lambda = Fa/k_BT$ and $\Gamma/\Lambda = \Omega/k_BT$. 
For this work, we will use these dimensionless parameters exclusively. 
In our study, we set temperature $\Lambda \in [0.1, 1.5]$, force $f \in [0,10]$, $\Gamma \in [-8,8]$,
and $N = 100$.
We find that this particular choice of $N$ is sufficient to capture relevant physics that govern our discretized polymer chain. 

\subsection*{Computational approach}
The ribbon is initialized with random orientations, with the one end of the chain fixed, 
namely, ${\bm F}_0 = (\mathbf{t}_0, \mathbf{n}_0, \mathbf{b}_0) = (\hat{\bf z}, \hat{\bf x}, \hat{\bf y})$, and the other chains free to take on arbitrary conformations. 
During the first step, we randomly picked two new angles $\theta_{1}$ and $\phi_{1}$, and update the adjacent Frenet frame ${\bm F}_1 = {\bm F}_{0} {\bm R}_1$ as well as the position of the next vertex, ${\bf x}_2$. This configuration is accepted via a Metropolis algorithm \cite{Sethna}. We then proceed to the next link and repeat the process. This procedure terminates when we reach the end of the chain and this constitutes one Monte Carlo sweep. For our simulations, we performed $10^{6}$ Monte Carlo sweeps per chain, the first half of which is devoted to equilibration. 
In our simulations, we have ignored the effects of knotting since the discrete chain segments are allowed to cross one another during trial moves in the Monte Carlo simulation.
We can close the open ribbon using the minimally-interfering closure scheme \cite{10.1143/PTPS.191.192}.
One would need to evaluate the Alexander polynomial for knot-checking and reject any trial moves that change the topology of the chain. 
Such a test is omitted in this study as it has been found that such an effect is not too significant, and topology checking is computationally intensive \cite{marko,moroz1,frank1}.

\section*{Ribbon under Tension} 

\subsection*{Force extension relations}

We first consider a ribbonlike polymer that is subject to an applied force without any external torque. 
A microscopic polymer chain behaves very differently from its macroscopic counterpart in that it is under constant thermal fluctuations, which prevents the chain from being straight.
Every Fourier mode of its shape is excited according to the equipartition theorem.
Because the ribbon is never straight, its average shape will respond as soon as any stress is applied. 
There is no threshold and it can bend and twist freely to relieve stresses.

\begin{figure*}[htb]
   \includegraphics[width=\textwidth]{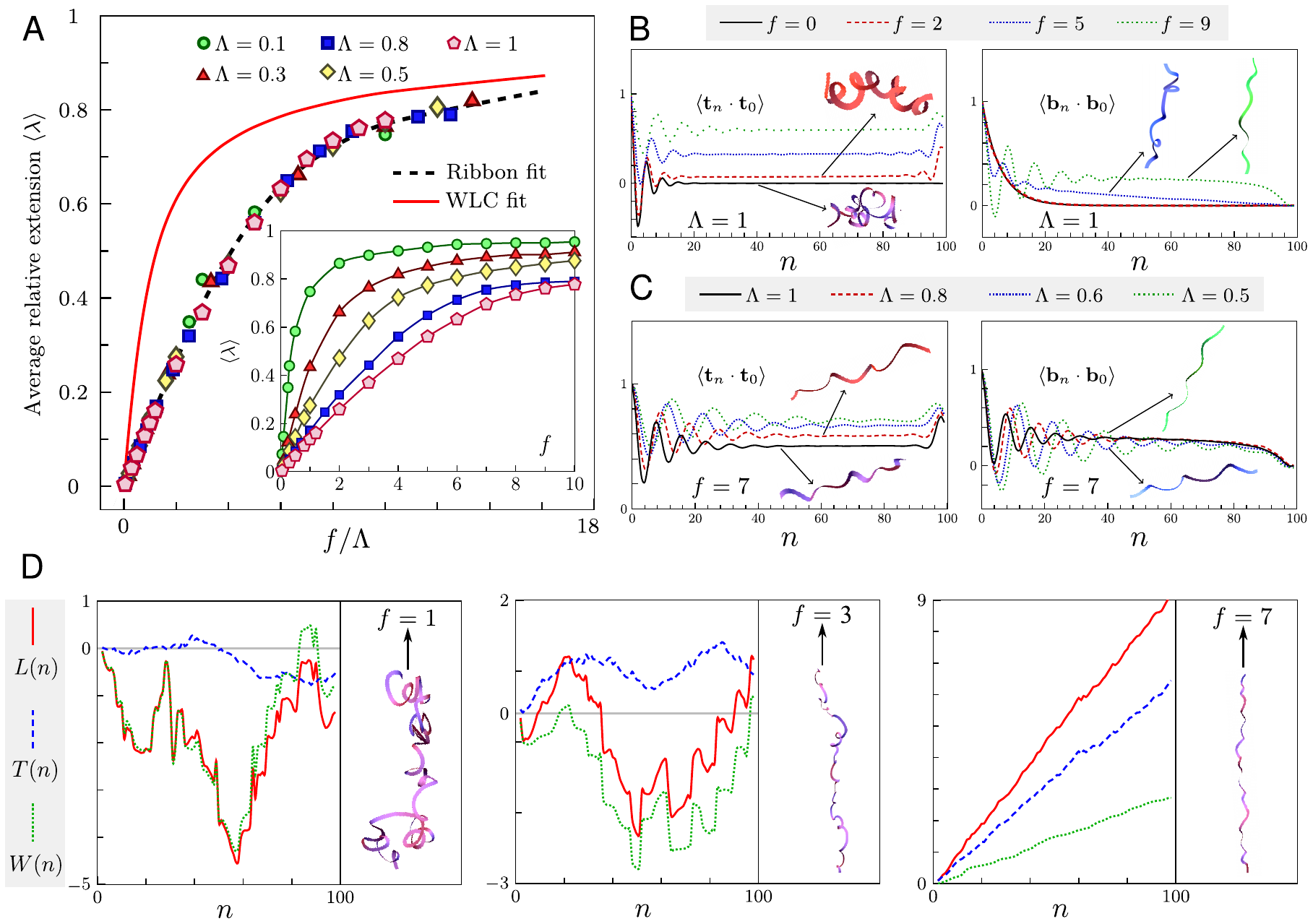}
   \caption{(A) The average relative extension $\langle\lambda\rangle$ for different force $f$ and temperature $\Lambda$ collapse onto a single curve (dashed line) fitted by Eq.~\ref{sadowfit}. 
   Inset:  The force-extension curves at different temperature, $\lambda$.
   (B) Tangent-tangent $\langle \mathbf{t}_{n}\cdot\mathbf{t}_{0}\rangle$ and binormal-binormal $\langle \mathbf{b}_{n}\cdot\mathbf{b}_{0}\rangle$ correlation functions from Monte Carlo simulations of a ribbonlike polymer chain of $N = 100$ segments at fixed temperature $\Lambda=1$ and different applied forces. 
   (C) $\langle \mathbf{t}_{n}\cdot\mathbf{t}_{0}\rangle$ and $\langle \mathbf{b}_{n}\cdot\mathbf{b}_{0}\rangle$ for $f=7$ ($f>f_c$) are plotted at different temperatures. (D) The cumulative density functions, $L(n)$, $T(n)$, and $W(n)$, along the ribbon for $f$ = 1 (HW phase), $f$ = 3 (HW phase), $f$ = 7 (HT phase) at $\Lambda$ = 1. The ribbon conformation is depicted beside each set of curves. The ribbon undergoes phase transition from HW to HT as we increase the force $f$. The critical force $f_c \approx 5$.
        }
        \label{Fig:notorque}
\end{figure*}

In the limit of low force ($f/\lambda <1$), the ribbon is in a random coil configuration and behaves like a linear spring with an elasticity that arises from entropy.
Fitting our simulations to the low force regime shows that the effective spring constant connecting the applied force $F$ to the relative extension $\lambda = z/L$ is given by $k \approx 7.15 k_BT/(La)$ as shown in Fig.~\ref{Fig:notorque}A. 
In terms of dimensionless parameters, this expression reads as
\begin{align}
\frac{f}{\Lambda} =\frac{Fa}{k_{B}T} \approx 7.15\frac{z}{L} = 7.15 \lambda.
\label{sadcurie} 
\end{align}
Evidently, the Sadowsky ribbon is stiffer than the WLC, which obeys $f/\Lambda = \frac{3}{2}\lambda$ in the limit of small force, thus requiring a much greater force to realize a given extension.
The ribbon becomes harder to stretch as the temperature $\Lambda$ increase.
As we increase the force, the force-extension curves becomes nonlinear. 
At higher forces ($f/\lambda > 10$) the relative extension begins to level off as it approaches unity. At this point the ribbon has been stretched nearly straight.
The relative extension $\lambda$ asymptotes toward $1$ with a distinctive $1/\sqrt{F}$ behavior.
In these regards, the Sadowsky ribbon is qualitatively similar to the WLC.
The force-extension curves at different temperatures collapse into one universal curve, as shown in Fig.~\ref{Fig:notorque}A.
The best-fit interpolation formulas for the WLC \cite{markosM} and Sadowsky ribbon as a function of $z/L$ are given by
\begin{align}
\frac{Fa}{k_{B}T} &= \frac{1}{4(1-\lambda)^{2}}+\lambda-\frac{1}{4} \quad \text{(WLC),}\label{WLCfit} \\
\frac{Fa}{k_{B}T} &= \frac{0.27}{(1-\lambda)^{2}}+6.60\lambda-0.27 \quad \text{(Ribbon)}.\label{sadowfit}
\end{align}


\subsection*{Geometrical correlations}

The tangent-tangent correlation function, $\langle\mathbf{t}_{n}\cdot \mathbf{t}_{m}\rangle$,  measures the correlations between the unit tangent at vertex $n$ and the unit tangent at vertex $m$; values close to 1 indicates high correlations while zero indicates no correlation. 
Since all the tangent vectors are equivalent, we compute $\langle\mathbf{t}_{n}\cdot \mathbf{t}_{0}\rangle$ for convenience. 
If the correlation length, $\ell$, which is the distance over which fluctuations in one region of space are correlated or affected by those in another region, is over multiple chain segments, the ribbon is said to exhibit long-range order; else, the ribbon is said to be disordered. 
Previous study \cite{lucamaha} has shown that the tangent-tangent correlation function, $\langle\mathbf{t}_{n}\cdot \mathbf{t}_{0}\rangle$, is oscillatory when $f=\Gamma=0$;
in fact, this is true for any finite $f$ and $\Lambda$ as shown in Fig.~\ref{Fig:notorque}B and \ref{Fig:notorque}C.
This means that the polymer model has an underlying long range ordered helical structure as long as there is no applied torque. 
For segments sufficiently far from the end segment, 
$\langle\mathbf{t}_{n}\cdot \mathbf{t}_{0}\rangle$ can be effectively described by 
\begin{align}
\langle\mathbf{t}_{n}\cdot \mathbf{t}_{0}\rangle = e^{-s/\ell_{p}}\cos(ks)+ M, 
\label{ttfiteq}
\end{align}
where $s=na$, $k$ is the wavenumber, $\ell_{p}$ is the persistence length characterizing the length scale over which orientational correlations persist, and $M$ is a constant. 
In the absence of external forces, the persistence length $\ell_p \sim a\Lambda^{-1}$ and the wavenumber $k \sim a^{-1}\Lambda^{1/2}$ \cite{lucamaha}. 
In addition, $M \to 0$ for $n \to \infty$ and the orientational correlation along the ribbon decays to zero. 
By contrast, for $f \ne 0$, $M \ne 0$ for $n \to \infty$, indicating long-ranged orientational order. 
The asymptotic value $M$, in turns, decreases as we increase the temperature, indicating a loss of long range order as shown in \ref{Fig:notorque}C. 
The tangent vector of the end segment $\mathbf{t}_N$, and its neighboring segments tend to align with the applied force ${\bm F} = F \hat{\bf z}$. Since $\mathbf{t}_0 = \hat{\bf z}$, this explains the peaks in tangent-tangent correlation function near the end segment.
On the other hand, the wavenumber $k$ at fixed $\Lambda$ gets smaller for a higher $f$, and eventually becomes smaller than the unit separation between each segments.

While $\langle\mathbf{t}_{n}\cdot \mathbf{t}_{0}\rangle$ is always oscillatory, we observe striking differences in behavior for the binormal-binormal correlation function $\langle\mathbf{b}_{n}\cdot \mathbf{b}_{0}\rangle$. 
When $f=\Gamma=0$, $\langle\mathbf{b}_{n}\cdot \mathbf{b}_{0}\rangle$ exhibits exponential decay at any nonzero temperature $\Lambda$; however, this is no longer true when $f$ is nonzero as plotted in Fig.~\ref{Fig:notorque}B and \ref{Fig:notorque}C.
As we increase $f$ at fixed $\Lambda$, we observe that $\langle\mathbf{b}_{n}\cdot \mathbf{b}_{0}\rangle$ experiences a transition at some critical force $f_{c} = f_c(\Lambda)$, going from (pure) exponential decay to oscillatory decay.
The binormal-binormal correlation function can be described as 
\begin{align}
\begin{split}
\langle\mathbf{b}_{n}\cdot \mathbf{b}_{0}\rangle=\begin{cases} 
      e^{-s/\ell_{\tau}}, & f<f_c(\Lambda)\\
      e^{-s/\ell_{\tau}}\cos({k_{\tau}s})+ M_{\tau}, & f>f_c(\Lambda)
   \end{cases}   
  \label{fittbbeq}
\end{split}
\end{align}
where $\ell_{\tau}$, $k_{\tau}$ and $M_{\tau}$ are the torsional persistence length, torsional wavenumber, and torsional LRO parameter, respectively.

\subsection*{Ribbon morphologies and chirality}

Even though the Sadowsky ribbon is achiral, 
individual torsional fluctuations will not be inversion symmetric, and different segments of the ribbon will exhibit different handedness. 
Two adjacent helical structures with opposite handedness are connected by a perversion, a classical motif commonly observed in tangled telephone cords and plant tendrils \cite{Keller, Darwin, Gerbode1087}. 
The ribbon will have segments of alternating chirality; 
along portions of the ribbon with right-handed (left-handed) helicity, $W(n)$ will increase (decrease) monotonically. 
Without perversions, the cumulative writhe density function can only change monotonically; at each perversion, the change in the cumulative writhe density function changes sign.
Under a small applied force ($f=1$, Fig.~\ref{Fig:notorque}D, leftmost panel), the ribbon is extremely coiled up, yet the twist and writhe are small, around $O(1)$ in magnitude. 
The fluctuations in the cumulative twist density function is significantly smaller than that in the cumulative writhe density function.
Through the proliferation of perversions, a ribbon that is bent and twisted at low applied force can achieve a small link. 
Because the link is not constrained by an external torque, it can be expelled from the free boundary \cite{Maha2019}.
We identify these kind of conformations as the helical phase that is writhe-dominated (HW), noting that it is the large change in cumulative writhe density function and the large numbers of perversions that captures the character of the ribbon conformations.
In the HW morphological phase at zero applied torque, the link, twist, and writhe are typically close to zero.
 
As we increase the applied force, the ribbon becomes increasingly stretched with a smaller number of perversions and the fluctuations in the cumulative twist density function become more significant.
At the critical force ($f=5$, Fig.~\ref{Fig:notorque}D, middle panel), the variations in cumulative twist density function become comparable to the cumulative writhe density function and we categorize the ribbon as simply in the helical state.
Under forces greater than $f_c$ ($f= 7$, Fig.~\ref{Fig:notorque}D, rightmost panel), the ribbon is nearly straight and has very few perversions. 
Simultaneously, the fluctuations in the cumulative link, twist, and writhe functions are increasingly suppressed and increase (decrease) monotonically if the helicity of the ribbon is right-handed (left-handed).
At zero torque, the ribbon can have either handedness as a result of the symmetry of the ribbon under parity. 
Here, we have used a right-handed conformation for illustration; in fact, the mirror image of this conformation is equally likely. 
In this case, we observe that Tw $\approx 6$, Wr $\approx 3$, and Lk $\approx 9$. This phenomenon is similar to the observation that the writhe in a helical telephone cord is converted into twist when stretched.  We term such morphologies as the helical phase that is twist-dominated (HT).  Thus, the Sadowsky ribbon at zero torque may exist either in the long range ordered twist-dominated helical phase (HT) where Tw $>$ Wr $\sim O(1)$ or the writhe-dominated helical phase (HW) where Wr $\approx$ Tw $\approx 0$, depending on $f$ and $\Lambda$. 
Generally, we observe that high/low force leads to twist/writhe dominated helical phase and
the transition point $f_c(\Lambda)$ occurs when the binormal-binormal correlation function changes from (pure) exponential decay to oscillatory decay. 
The cumulative density functions allow us to investigate the formation of perversions as we move along the arc length of the ribbon. 
In our model, the morphological crossover is also accompanied by a spontaneous breaking of the symmetry of the ribbon under parity, which will be discussed later in the next section.


\section*{Ribbon under Tension and Torque}
     
\subsection*{Force extension relations under torque}

\begin{figure*}[htbp!]
   \includegraphics[width=\textwidth]{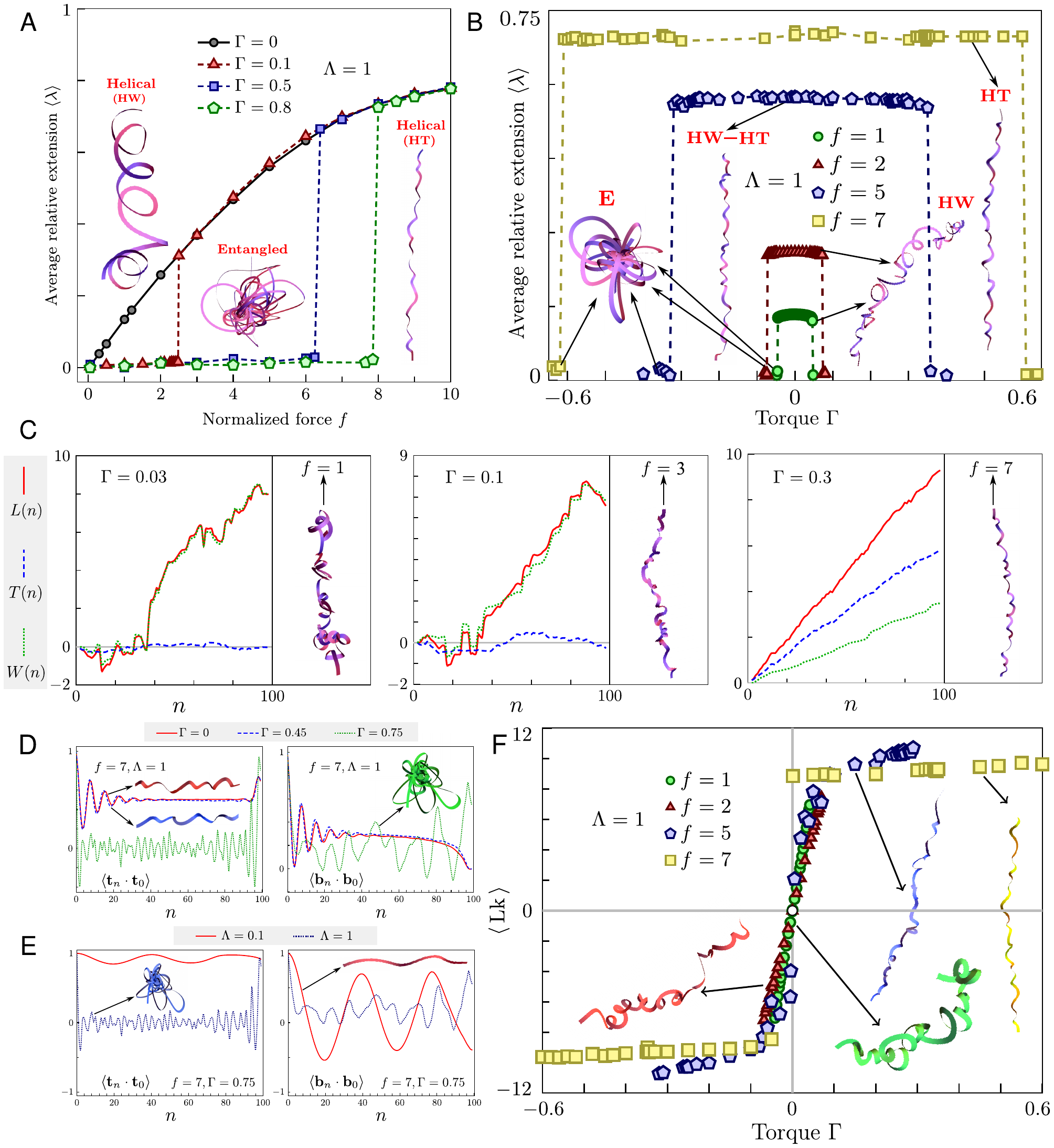}
      \caption{
(A) The force-extension curve for a ribbon at temperature $\Lambda = 1$ under different applied torque $\Gamma$. 
(B) Average relative extension $\langle\lambda\rangle$ vs applied torque $\Gamma$ of a ribbon at fixed temperature for different force. The helical phase is stable for $\Gamma \in [-\Gamma_{c}(f, \Lambda), \Gamma_{c}(f, \Lambda)]$; otherwise the ribbon is entangled. 
(C) The cumulative density functions, $L(n)$, $T(n)$, and $W(n)$, along the ribbon for $f = 1$ and $\Gamma =0.03$ (HW phase), $f = 3$ and $\Gamma =0.1$ (HW phase), and $f$ = 7 and $\Gamma = 0.3$ (HT phase) at $\Lambda$ = 1. 
(D) Tangent-tangent and binormal-binormal correlation functions corresponding to $f=7$ and $\Lambda=1$ for different torque $\Gamma$. The phase transition occurs at $\Gamma_{c}\approx 0.62$. 
(E) $\langle \mathbf{t}_{n}\cdot\mathbf{t}_{0}\rangle$ and $\langle \mathbf{b}_{n}\cdot\mathbf{b}_{0}\rangle$ for $f=7$ ($f>f_c$) and $\Gamma = 0.75$ are plotted at different temperatures. 
(F) Plot of average link vs torque at $\Lambda = 1$. For $f < f_c$, the average link varies continuously with applied torque. At critical force $f_c = 5$, the slope approaches infinity. For $f> f_c$, the average link is discontinuous and the jump in $\langle \text{Lk} \rangle$ at zero applied torque is indicative of a first-order phase transition. At zero torque, parity symmetry is spontaneously broken at large $f$ and the ribbon adopts a particular handedness.
}     
      \label{Fig:torque}
\end{figure*} 

A macroscopic elastic rod under torque and tension remains straight as long as the twisting couple is below a critical value 
$\Omega_{\text{critical}} \propto \sqrt{F}$ 
\cite{love, moroz1,moroz2}. 
On the other hand, a microscopic ribbon under tension and torque behaves very differently as there is no threshold to bending or twisting.
In this case, the twisting couple term breaks the parity symmetry of the Sadowsky functional:
$\Omega > 0$ will lead to a right-handed ribbon while $\Omega <0$ will result in a left-handed ribbon.
The twisting couple will push fluctuations with the same handedness closer to instability, while suppressing those of the opposite handedness, 
and ribbon conformations with the same helicity as the applied torque will be favored.
These twist fluctuations will in turn affect the bend fluctuations due to the coupling between the curvature and torsional modes of the Sadowsky ribbon. 
The end result will be a coupling between the applied torque and the mean end-to-end extension of the polymer. 

The force-extension 
curve of our polymer chain is shown in Fig. \ref{Fig:torque}A together with characteristic conformations of the ribbon for $\Lambda=1$, and different $\Gamma$.   
Under small applied torques, the ribbon is in the HT phase at high force.
As we reduce the force, there is an crossover from twist to writhe, and the ribbon typically becomes shorter and more coiled and transit from HT phase to HW phase.
Under zero torque, the ribbon will shift from HT to HW phase under decreasing applied force;
for nonzero torque, 
the ribbon rapidly goes into
an entangled phase (E) below a minimum force $f_c(\Gamma, \Lambda)$. 
This transition into entangled phase may occur either from the HT or HW phase depending on the force, torque, and temperature.
The entangled morphological phase is characterized by its small relative extension ($\lambda \ll 1$).
As we do not include steric effects, the ribbon in the entangled phase collapses into a small globule. Due to the self-crossing, the topology of the ribbon will be heavily knotted. 
The helix to entangled transition is abrupt, reminiscent of a first-order phase transition and 
critical force becomes smaller with either decreasing torque or temperature.

The average relative extension of the ribbon $\langle \lambda \rangle$ under varying torque $\Gamma$ for different values of $f$ and $\Lambda$ is shown in Fig.~\ref{Fig:torque}B. 
Due to the assumption of developability, 
the relative extension is an even function of the applied torque. 
A ribbon under high applied force (yellow squares, Fig.~\ref{Fig:torque}B) is stable in the HT phase as long as
$\Gamma < |\Gamma_{c}(f, \Lambda)|$; otherwise, it becomes entangled.
In the HT phase, increasing the torque does not result in significant changes to the ribbon conformations. 
The change in relative extension whilst in the helical phase is relatively small, $\Delta\lambda = \lambda_{\text{max}}-\lambda_{\text{min}} \approx 0.02$. 
As we reduce the applied force, the helical phase remains stable for a smaller range of applied torque, i.e. $\Gamma_{c}(f, \Lambda)$ reduces with decreasing applied force.

\subsection*{Statistical topology of ribbons}

Under a small applied force and torque ($f=1$, $\Gamma = 0.03$, Fig.~\ref{Fig:torque}C, leftmost panel), the ribbon is coiled in a random manner. 
We observe that the cumulative writhe density function is significantly larger than the cumulative twist density function and the ribbon is evidently in the HW phase.
Due to the symmetry breaking torsional constraint term, right-handed fluctuations will be favored:
ribbon segments with right handed chirality tend to be extended over many discrete segments while those with left-handed chirality are generally short ranged.
The overall chirality of the ribbon is right-handed, resulting in a non-zero, positive link in this case. 
The twist of the ribbon is close to zero and the writhe provides the leading contribution to the link of the ribbon.
A similar behavior is observed at $f=3$ and $\Gamma = 0.1$ (HW phase), as observed in the middle panel of Fig.~\ref{Fig:torque}C. 
Unlike the zero torque case considered in the previous section, the HW phase under torque has a finite writhe, i.e. Wr $\sim O(1)$ and Tw $\approx 0$. 
As we increase the applied force and torque ($f=7$, $\Gamma = 0.3$), the ribbon becomes elongated and transition into the HT phase as seen in rightmost panel of Fig.~\ref{Fig:torque}C, and Tw $>$ Wr $\sim O(1)$.
We observe the disappearance of perversions and the cumulative density functions are generally monotonic. 
The ribbon has a right-handed helicity, with the twist as the predominated contributor to the link in this case.

\subsection*{Geometrical correlations}
Whilst in the helical phase (HW/HT), 
the correlation functions $\langle\mathbf{t}_{n}\cdot \mathbf{t}_{0}\rangle$ and $\langle\mathbf{b}_{n}\cdot \mathbf{b}_{0}\rangle$ obey Eq.~\ref{ttfiteq} and \ref{fittbbeq} respectively;
in the entangled phase, $\langle\mathbf{t}_{n}\cdot \mathbf{t}_{0}\rangle$ and $\langle\mathbf{b}_{n}\cdot \mathbf{b}_{0}\rangle$ exhibit random oscillatory behaviors as shown in Fig.~\ref{Fig:torque}D; the transition occurs precisely at $\Gamma_c(f, \Lambda)$.
The critical point that divides the disorder entangled phase and the ordered helical phase is traditionally referred to as the ``Lifshitz point'' \cite{liv,lucamaha}. 
The transition may be achieved either by varying the force at fixed temperature and torque $f_c(\Gamma, \Lambda)$, changing the torque at fixed temperature and force $\Gamma_c(f, \Lambda)$ (Fig.~\ref{Fig:torque}D), or changing the temperature at fixed force and torque $\Lambda_c(f, \Gamma)$ (Fig.~\ref{Fig:torque}E).

In the presence of force, we observe that both $\ell_p$ and $\ell_\tau$ are approximately constant in the HW phase and and scale as $f^{-1/2}$ in the HT phase. Interestingly, the high $f$ limit is similar to that of the Moroz-Nelson (MN) model \cite{moroz1}.
We find that $k$ scales as $(f^{2}+C)^{-1}$ over all $f$, where $C$ is a temperature-dependent constant that keeps $k$ finite in the low-force limit; $k_\tau$ vanishes in HW phase and scales as $f^{-2}$ in the HT phase. 
The temperature dependence at finite $f$ appears to be identical to the $f=0$ case: both $\ell_p$ and $\ell_\tau$ scale as $\Lambda^{-1}$, 
$k$ scales as $a\Lambda^{1/2}$ in either helical phase, and $k_\tau$ scales as $a^{-1}\Lambda^{1/2}$ in HT phase.
Finally, we observe that $\ell_p$, $\ell_\tau$, $k$, and $k_\tau$ do not seem to vary appreciably with $\Gamma$ in the range investigated in this work. 

\subsection*{Phase transition with Lk as the order parameter}

The variation of $\langle\text{Lk}\rangle$ of the ribbon in the helical phase at fixed temperature with the applied torque $\Gamma$ is shown in Fig.~\ref{Fig:torque}F. 
At forces below the critical force $f_c$, the ribbon has zero average link at zero torque. 
The link exhibits a linear dependence on the applied torque. 
At forces above $f_c$, there is a jump in $\langle \text{Lk}\rangle$ at zero applied torque, indicative of a first-order phase transition, i.e. $\langle \text{Lk}\rangle = \pm \text{Lk}_0$ at $\Gamma =0$.
The link changes slightly with the applied torque and asymptotes to a constant that represents the maximum link admissible to the chain with $N$ finite segments in helical phase.
 At $f_c$, $\langle \text{Lk}\rangle$ is continuous at $\Gamma=0$ but has an infinite slope. 
This is analogous to what happens in ferromagnets: the magnetization varies discontinuously with the applied field when the temperature is below the Curie temperature $T_{\text{C}}$ (ferromagnetic phase), while the magnetization varies smoothly when $T> T_{\text{C}}$ (paramagnetic phase).
Thus, we can interpret the average link $\langle\text{Lk}\rangle$ as the order parameter of the ribbonlike polymer and the torque $\Gamma$ as the conjugate field. 
In the HW morphological phase at zero torque, the ground state of the ribbon has zero link, i.e. $\langle\text{Lk}\rangle = 0$, and the system exhibits chirality. 
However, in the HT phase at zero torque, the ground state becomes doubly degenerate, and
the ribbon commits to one of the two minima ($\pm$ $\text{Lk}_0$) randomly, resulting in the spontaneous breaking of chirality. 
Unlike in the simple ferromagnet where the transition is controlled by the strength of external field, the morphological transition of the ribbonlike polymer can be achieved by either tuning temperature, force, or torque. 

\label{conclusion}

\begin{figure}[htb]
   \includegraphics[width=0.5\textwidth]{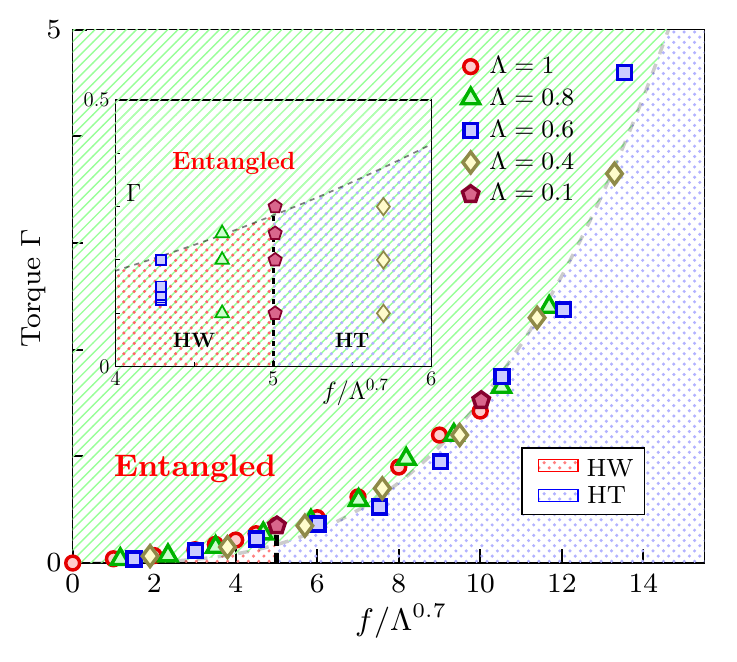}
   \caption{Morphological phase diagram of the ribbonlike polymer under different torque, force, and temperature. Three morphological phases exist: writhe-dominated helical (HW), twist-dominated helical (HT), and entangled (E). Transition between HW and HT is continuous and phase boundary occurs at $f\approx5\Lambda^{0.7}$. Transition between HW/HT and E is abrupt and the phase boundary collapses onto a universal curve $\Gamma_{c}(f, \Lambda)=g(f/\Lambda^{0.7})$ where $g(x)\sim x^3$.
   }
    \label{Fig:Phase}
\end{figure}

\subsection*{Morphological phase diagram}

Our study shows that the Sadowsky ribbon exhibits a rich morphological phase diagram that span the torque-force-temperature phase space as shown in Fig.~\ref{Fig:Phase}.
There are three phases in the ribbonlike polymer: writhe-dominated helical phase (HW) at low force and low torque, twist-dominated helical phase (HT) at high force and low torque, and the entangled phase at high torque. 
Both helical phases are ordered while the entangled phase is disordered. 
At zero torque, the ribbon conforms to either HW or HT phase at all finite $f$ and $\Lambda$.
In the HW phase at zero torque, the ground state of the ribbon is chiral with zero average link with multiple perversions while in the HT phase, the ribbon relaxes into either one of two minima ($\pm$ Lk$_0$) and parity symmetry is spontaneously broken. 
The transition occurs at $f_c(\Lambda)$ whereby the binormal-binormal correlation function changes from (pure) exponential decay to oscillatory decay. 

At finite torque, the HW phase is characterized by its small relative extension, small average link, tangent-tangent correlation function that is oscillatory, and binormal-binormal correlation function that is purely exponential decay. 
The link is shown to increase with the applied stresses and the ribbon has finite writhe and small twist. 
There are multiple perversions that flips the chirality of the ribbon and its overall handedness conforms to that of the applied torque since fluctuations with the same handedness are favored.
In the HT phase, the ribbonlike polymer has large relative extension ($\langle \lambda \rangle \to 1^-$), finite $\langle \text{Lk} \rangle$ (Tw $>$ Wr), and oscillating correlation functions.
The ribbon tend to be straight as a consequence of the nonlinear coupling term $\tau^{4}/\kappa^{2}$ which forces $\tau$ to approach zero faster than $\kappa$. 
The ribbon has few if any perversions and the link changes marginally from $\pm$ Lk$_0$  under applied stresses. 
The ribbonlike polymer in HW/HT phase experiences very small variations in its relative extension under variation in torque.
The transition between HW and HT phase is continuous and phase boundary occurs at $f\approx5\Lambda^{0.7}$, independent of the applied torque.
It is stable until a critical torque $\Gamma_{c}(f,\Lambda)$, beyond which the ribbon starts to get entangled, reminiscent of a first-order phase transition.
In the entangled phase, the polymer has extremely small $\langle \lambda \rangle$, diverging $\langle \text{Lk} \rangle$, and correlation functions with random oscillations.
The transition between HW/HT and E phase is abrupt and the phase boundary collapses onto a universal curve $\Gamma_{c}(f, \Lambda)=g(f/\Lambda^{0.7})$ for $0\leq f/\Lambda^{0.7}\leq 16$, where $g(x) \sim x^3$ is a function of one variable. The data was fitted to $Ax^{B}$, with $A=0.0011\pm 0.0002$, $B = 3.14\pm 0.08$, and adjusted R-Squared $\approx 0.98$.

\section*{Conclusions}

In this work, we combine statistical mechanics, elasticity, topology, and geometry to explain the complex morphology of ribbonlike polymers whose dimensions are well separated in length scale. 
The ribbon can exist in three morphological phases: writhe-dominated helical, twist-dominated helical, and entangled depending on the applied tension, torque, and temperature. 
We characterize the rich set of ribbon conformations using cumulative density functions and explain how the link, twist, and writhe are related to its geometry and shape. 
Our study has a set of clear experimental predictions for the equilibrium statistical mechanics of
inelastic thermal ribbon under tensile and torsional stress.

We can use existing code \cite{Tubiana2018} to identify the knot types
and answer question such as: Is there any connection between topology and morphology? 
Additionally, in our study, we have shown that perversions arise naturally in the HW phase but there are several open questions on that remain unanswered: How often do perversions appear in the HW phase? How far apart are the perversions? Where are the perversions located? 
Lastly, we do not see more complicated objects such as plectonemes and solenoids in our simulations. How do we modify the energy function (Eq.~\ref{Eq:TotalE}) to add additional realism to this ribbon model? 
These are important topics of future study.


%
%




E.H.Y. and L.M. conceived and designed research; E.H.Y. and F.D. performed research; E.H.Y., F.D., L.G., and L.M. analyzed data; and E.H.Y., F.D. and L.M. wrote the paper. The authors declare no competing interest. E.H.Y. and F.D. contributed equally to this work.


E.H.Y. and F.D acknowledges support from Nanyang Technological University, Singapore, under its Start Up Grant Scheme (04INS000175C230). The computational work for this article was performed on resources of the National Supercomputing Centre, Singapore (https://www.nscc.sg). L.G. is supported by the European Union via the ERC-CoG grant HexaTissue and by Netherlands Organization for Scientific Research (NWO/OCW), as part of the Vidi scheme. LM thanks the NSF Harvard MRSEC DMR 2011754, NSF EFRI 1830901, the Simons Foundation and the Seydoux Fund for partial financial support. 


\bibliography{references}

\end{document}